\documentclass[a4paper,aps,pra,showpacs,superscriptaddress,twocolumn,nofootinbib]{revtex4-1}

\usepackage{graphicx,graphics,epsfig}   
\usepackage{dcolumn}    
\usepackage{bm}         
\usepackage{amsmath}    
\usepackage{verbatim}   
\usepackage{color}      
\usepackage{subfigure}  
\usepackage{times,natbib}
\usepackage{amsmath,amsfonts,amssymb,graphics,graphics,color,times}

\usepackage{latexsym}
\usepackage{amsmath}
\usepackage{amssymb}
\usepackage{amsfonts}
\usepackage{amsthm}
\usepackage{mathrsfs}
\usepackage{color,verbatim,graphics}
\usepackage{psfrag}
\DeclareMathAlphabet{\mathrsfs}{U}{rsfs}{m}{n}
\DeclareMathAlphabet{\mathpzc}{OT1}{pzc}{m}{it}
\DeclareMathAlphabet{\matheus}{U}{eus}{m}{n}
\DeclareMathAlphabet{\mathbbold}{U}{bbold}{m}{n}

\def\one{\leavevmode\hbox{\small1\normalsize\kern-.33em1}}

\newcommand{\ba}{\begin{eqnarray}}
\newcommand{\ea}{\end{eqnarray}}
\newcommand{\ban}{\begin{eqnarray*}}
\newcommand{\ean}{\end{eqnarray*}}

\newcommand{\etal}{{\it{et al.~}}}

\begin{document}

\title{Bell inequalities violated using detectors of low efficiency}

\author{K\'aroly F. P\'al}

\author{Tam\'as V\'ertesi}
\affiliation{Institute for Nuclear Research, Hungarian Academy of Sciences, H-4001 Debrecen, P.O. Box 51, Hungary}

\date{\today}


\begin{abstract}
We define a family of binary outcome $n$-party $m\leq n$ settings per party Bell inequalities whose members require the least detection efficiency for their violation among all known inequalities of the same type. This gives upper bounds for the minimum value of the critical efficiency --- below which no violation is possible --- achievable for such inequalities. For $m=2$, our family reduces to the one given by Larsson and Semitecolos in 2001. For $m>2$, a gap remains between these bounds and the best lower bounds. The violating state near the threshold efficiency always approaches a product state of $n$ qubits.
\end{abstract}

\pacs{03.65.Ud, 03.67.-a}

\maketitle

\section{Introduction}

In a Bell experiment distant parties perform measurements on a shared physical system. From very reasonable assumptions it follows that the correlations between their results must satisfy certain inequalities. Quantum mechanics predicts that often these Bell inequalities will not be respected \cite{Bell,review}. Their violation, which excludes all local realistic explanations for the quantum world, and which is often referred to as nonlocality, is one of the most surprising and counterintuitive features of quantum physics. One of the earliest experimental demonstrations of the Bell violation were done by Aspect \etal \cite{aspect}, and many experiments on various systems have been performed since (see, e.g.\ Ref. \cite{Bellexper}). Unfortunately, due to technical imperfections there are still loopholes, which in principle allow local realistic explanation for the experimental results.

An important loophole, the so-called detection loophole appears in experiments performed on systems of photons~\cite{pearle}. Some photons get lost during transmission and the efficiency of the present day photon detectors is also limited. If the proportion of photons detected falls below a certain critical value $\eta_{crit}$, which depends both on the Bell inequality and on the system, then the results become compatible with local realism. As an important property characterizing the inequality and the system, Vallone \etal \cite{Vallonetal} introduced the notion of the robustness of nonlocality $R$. They defined this quantity as the maximum fraction of detection events that can be lost such that the remaining ones still do not admit a local model. It is easy to see that $R=1-\eta_{crit}$.
In the present paper it will be assumed that the detection efficiency is the same for all parties. There are experiments, for example those involving entangled photon--ion systems, where some of the parties (the ones working with the ions) have virtually no loss of events. Then the experiment tolerates higher losses for the remaining parties than in the symmetric case considered here~\cite{asymsetup}.

Another important loophole is the locality loophole, whose closure requires space-like separation between the parties. This can realistically be ensured only in Bell tests performed on photon systems. Each loophole has been closed by some experiment, but not all simultaneously, for example, no single experiment closed the locality and the detection loophole at the same time (for a comprehensive review, see Ref.~\cite{larsson_review}). Such a verification would be interesting and fairly important from a philosophical point of view, although  --- as so many predictions of quantum mechanics has proven correct --- at present virtually everyone is quite convinced that the Bell violations observed so far are genuine, and not results of some conspiracy plotted by nature, using different loopholes in different situations to mislead us.

However, there are also practical reasons why loophole free Bell violation would be important to achieve. Bell violation is the basis of the so-called device independent protocols of quantum information technology (see for example Ref.~\cite{scarani} for a recent review). Such protocols, based on the fact that Bell violation is very hard to fake, would make it possible for anyone to check the proper functioning of an apparatus without knowing much about the details of its internal structure. Therefore, construction faults would be easier to notice and it would be harder for a malicious manufacturer to build back doors into the device. This approach promises to perform cryptographic tasks with an unprecedented security~\cite{key}, produce genuinely random numbers~\cite{randomness}, and carry out black-box state tomography~\cite{tomography}.

The violation of the simplest Bell inequality, the two-party two-setting per party ($n=m=2$) Clauser-Horne-Shimony-Holt (CHSH) inequality~\cite{chsh} requires at least $\eta_{crit}=2/3\approx 0.667$ detection efficiency \cite{eberhard}. This value may be reached with a partially entangled two qubit state, which approaches a product state near the threshold efficiency. In this case the robustness of nonlocality is $R=1/3$. As far as we know, with two parties performing measurements on a pair of qubits no better value has been achieved. In Ref.\cite{VPB} a bipartite four-setting inequality was found with a somewhat lower critical efficiency of $\eta_{crit}=(\sqrt{5}-1)/2\approx 0.618$. To get this value a pair of ququarts had to be used. Larsson and Semitecolos~\cite{LS} have introduced a family of $n\geq 2$ party binary Bell inequalities with $m=2$ settings per party with $\eta_{crit}=n/(2n-1)$. They have also proved that this value is optimal, no two-setting $n$-party Bell inequality may be violated with any lower detection efficiency. This means that for $m=2$ an efficiency of less than $1/2$ is never enough.

Let $\eta_*$ be the smallest of the critical detection efficiencies belonging to the Bell inequalities characterized by a certain number of parties and numbers of settings and measurement outcomes for each party. Above $\eta_*$ there are quantum measurements exhibiting correlations that can not be explained by any local hidden variables model. The value for $\eta_{crit}$ given by Ref.~\cite{LS} and cited above is just $\eta_*$ for binary-outcome n-party two-setting per party Bell inequalities. For $n$-party and more than two-setting per party ($m>2$) inequalities $\eta_*$ is not known.
Massar and Pironio \cite{massar} has given a lower bound for this quantity by constructing explicit local hidden variables models reproducing the correlations for efficiencies below this bound.
For $m=2$ the bound is the same as the one of Ref.~\cite{LS}. Based on combinatorial considerations Buhrman \etal \cite{BHMR} has derived an upper bound for $\eta_*$, when $m$ is some power of two. Ref.~\cite{PVB} also provided upper bounds for many ($n$,$m$) combinations ($m$ prime) by giving explicit Bell inequalities corresponding to those efficiency values. The inequalities of that work may become useful in practice as well, because they require the Greenberger--Horne--Zeilinger (GHZ) state \cite{ghz}, very well distinguishable measurement settings even near the threshold efficiency, and the noise tolerance is also quite good.

Very recently tripartite Bell inequalities have been constructed~\cite{PVw} requiring low detection efficiencies for their violation with the $W$ state~\cite{dur}. With three measurement settings for one of the parties and two settings for the other two parties a critical efficiency of $\eta_{crit}=0.6$ has been achieved. The same efficiency has been necessary for an inequality symmetric for the permutations of the parties with $m=3$ settings per party. For $m=4$, $6$ and  $8$ this value has been improved to $\eta_{crit}=0.509036$, $0.502417$ and $0.501338$, respectively. For larger $m$ the construction would have required too much computational resources but the result would probably have remained above $0.5$. However, if a small admixture of a product state to the $W$ state has been allowed, $\eta_{crit}=0.5$ has been found for an $m=4$ inequality. This value could not be improved by increasing $m$.

In this work, we have used an iterative procedure alternating a
linear programming step and a semidefinite programming one to get
the $m=3$, $n=3$ and $n=4$ Bell inequalities with the smallest
possible $\eta_{crit}$. In this procedure we made assumption
neither about the symmetry of the inequality nor about the
properties of the state, not even about the dimensionality of the
Hilbert space. We have got tripartite inequalities reaching
$\eta_{crit}=0.5$ already with three settings per party, and for
$m=3$, $n=4$ we have obtained $\eta_{crit}=6/13\approx 0.46154$.
We could get the same values with confining ourselves to
permutationally symmetric inequalities. Numerical calculations
showed that the state giving the maximum violation near the
threshold efficiency is a state of $n$ qubits, namely a mixture of
the $W$ state and a product state that approaches the product
state near $\eta_{crit}$, the same behavior as the one observed for
the two-setting inequalities of Larsson and Semitecolos~\cite{LS}.
The measurement settings could be chosen the same for all parties.
Each measurement operator could be characterized by a single angle
that approached zero at $\eta_{crit}$, but at different paces for
the different settings, again similarly to what is seen in
Ref.~\cite{LS}. These observations helped us to make the
generalization and to define a whole family of Bell inequalities
with any $m$ and $n\ge m$. The members of the family require
the least detection efficiency for their violation among all known
inequalities of the same type, giving upper bounds for $\eta_*$.
For $m=2$ the inequalities are the same (apart from swapping of the two measurement settings) as the ones in Ref.~\cite{LS}.

The structure of this paper is as follows. In Section~\ref{ineq},
we give explicitly the family of multipartite Bell expressions.
Then, in Section~\ref{qviol}, we give the optimal quantum
violation of these Bell inequalities with qubit systems accounting
for finite detection efficiencies. Our main result is presented in
formula~(\ref{eq:etacrit}) providing (to the best of our
knowledge) the best upper bound on $\eta_*$ for any $n\ge m$
parties and $m$ settings per party. To get this formula, we have
used two claims: one about the equality of certain
quantum conditional probabilities and the negligibility of others, while the other concerning the classical bound. 
The first claim is proven in Section~\ref{qcond}, while the second one in
Section~\ref{class}. To demonstrate the behavior of the relevant quantities, we
highlight the example $n=m=3$ in section~\ref{333}. Here, we give
the explicit Bell expression and also show the optimal measurement
settings (angles) and Bell violation departing from the value of
the critical detection efficiency. Finally, we summarize our
conclusions.

\section{The inequalities}\label{ineq}
We consider the Bell scenario with $n$ distant observers. Each
observer may freely choose between a set of $m$ binary measurement
settings with possible outcomes zero and one. Let $A_{ij}$ denote
the $j$th measurement of the $i$th party. A Bell inequality
corresponding to this setup may be written as:
\begin{equation}
\sum_{j_1=0}^m\sum_{j_2=0}^m\dots\sum_{j_n=0}^m  B_{j_1j_2\dots j_n}
P(A_{1j_1}A_{2j_2}\dots A_{nj_n})\leq L,
\label{eq:Bellineq}
\end{equation}
where $L$ is the classical bound, and
$P(A_{1j_1}A_{2j_2}\dots A_{nj_n})\equiv P(11\dots 1|A_{1j_1}A_{2j_2}\dots A_{nj_n})$
denotes the conditional probability of all parties getting outcome one given they have measured
$A_{1j_1}$, $A_{2j_2}$, and so on. For the sake of brevity of the notation, for terms involving only a subset of observers, we have introduced a zeroth measurement whose outcome is always one, and which is always supposed to be performed. This measurement appears in these terms for the observers not members of the subset considered.

The inequalities we introduce in the present paper are invariant with respect to the permutations of the parties, and $n\geq m$ holds. Besides terms involving all $n$ observers, they contain only terms with $n-1$ parties (that is terms containing equal or less than $n-2$ parties are zero). Due to the permutational symmetry, Bell coefficients having the same set of indices in different orders have the same value. Therefore, we only give explicit values for coefficients with indices $j_1\leq j_2\leq j_3\dots\leq j_n$.

The nonzero $(n-1)$-party Bell coefficients have the value of minus one. Besides the index zero for the party not involved, they have indices from 2 to $m$ once, plus index $m$ another $n-m$ times if $n>m$. For example, if $m=2$ and $n>2$ then $B_{02\dots 2}=-1$, or if $m=3$ and $n>3$ then $B_{023\dots 3}=-1$. For larger values of $m$ the nonzero coefficients are
\begin{equation}
B_{0234\dots mm\dots m}=-1,
\label{eq:nm1partycoeff}
\end{equation}
and the ones we get by permuting the indices of the coefficients above. Thus there are $n!/(n-m+1)!$ nonzero $(n-1)$-party coefficients.

Furthermore, we have $n$-party Bell coefficients whose value is positive. They are the ones that have the same set of indices as the nonzero $(n-1)$-party coefficients above, but index zero is replaced by anything from one to $m$. If this index is $m$, the value of the coefficient is $(n-m+1)$, otherwise it is one. Therefore, we may write (again for larger $m$)
\begin{align}
B_{1234\dots mm\dots m}&=1\nonumber\\
B_{2234\dots mm\dots m}&=1\nonumber\\
B_{2334\dots mm\dots m}&=1\nonumber\\
B_{2344\dots mm\dots m}&=1\nonumber\\
\vdots\quad\quad\quad\quad\quad&\nonumber\\
B_{234\dots mmm\dots m}&=n-m+1.
\label{eq:npartycoeff}
\end{align}
Again, all coefficients we get by permuting the indices have the same value. Thus the first and the last line represent $n!/(n-m+1)!$ and $n!/(n-m+2)!$ coefficients, respectively, while the lines in between represent $n!/2(n-m+1)!$ coefficients.

There are also some $n$-party Bell coefficients having negative values:
\begin{equation}
B_{k_1k_2k_3k_4\dots k_n}<0,
\label{eq:npartycoeffneg}
\end{equation}
where the indices are such that the quantum values of the conditional probabilities multiplying these coefficients calculated with the quantum state and the measurement operators we will
give later are vanishingly small compared to the quantum conditional probabilities
multiplying the coefficients given by Eqs.~(\ref{eq:nm1partycoeff}) and (\ref{eq:npartycoeff}).
Then the actual values of these coefficients will not influence the quantum value of the Bell expression (the left hand side of Eq.~(\ref{eq:Bellineq}), therefore we may choose them freely.
As we will show later, this freedom makes it possible to achieve that the classical bound $L$ appearing in Eq.~(\ref{eq:Bellineq}) is zero, which is appropriate for our purpose. There are many solutions, and they all give the same value for the threshold detector efficiency. The maximum quantum violation above the threshold efficiency does depend on the actual choice, but we have made no attempt to find the best one.

All other Bell coefficients are zero.

\section{Quantum violation with detectors of limited
efficiency}\label{qviol}
Let us consider the quantum violation of
our Bell inequality. If we confine ourselves to von Neumann
measurements on a pure $n$-qubit state $|\psi\rangle$, the quantum
value of the conditional probability appearing in
Eq.~(\ref{eq:Bellineq}) may be written as:
\begin{equation}
P_Q(A_{1j_1}A_{2j_2}\dots A_{nj_n})=\langle\psi|\bigotimes_{i=1}^n\hat A_{ij_i}|\psi\rangle,
\label{eq:qvalprob}
\end{equation}
where $\hat A_{ij}$ for $1\leq j\leq m$ is the measurement operator corresponding to measurement $A_{ij}$, $\hat A_{i0}=\hat I$ is the identity operator in the subspace of the  observer, and $\bigotimes_{i=1}^n\hat A_{ij_i}$ denotes the tensor product of the operators of the parties. The state $|\psi\rangle$ we consider in this paper is invariant with respect to the permutations of the parties, and all parties have the same set of measurement operators, that is $\hat A_{ij}=\hat A_{j}$ is independent of $i$. Then, if the Bell inequality of
Eq.~(\ref{eq:Bellineq}) is permutationally invariant and the classical bound is zero, the condition for its quantum violation can be written as:
\begin{align}
&\sum_{j_1=0}^m\sum_{j_2=j_1}^m\dots\sum_{j_n=j_{n-1}}^m\pi(j_1,j_2,\dots,j_n)B_{j_1j_2\dots j_n}\nonumber\\
&\langle\psi|\bigotimes_{i=1}^n\hat A_{j_i}|\psi\rangle>0,
\label{eq:Bellqviol}
\end{align}
where $\pi(j_1,j_2,\dots,j_n)$ denotes the number of independent permutations of its arguments.
Let of suppose that the observers have detectors of limited efficiency $\eta<1$. Let they agree that each of them signal outcome zero whenever the particle is not detected. In this case each term in the left hand side of Eq.~(\ref{eq:Bellqviol}) must be multiplied by the probability of detecting all particles concerned to get the condition for the detection of the violation. This probability is $\eta^l$ for an $l$-party term, where $l$ is the number of nonzero indices. At the threshold efficiency $\eta_{crit}$ the resulting expression is zero. This condition gives an $n$th order equation for $\eta_{crit}$. In our case the solution is simple, because we have restricted ourselves to $n$-party and $(n-1)$-party terms. The condition for being able to detect the violation is
\begin{align}
&\eta^{n-1}\sum_{j_2=1}^m\sum_{j_3=j_2}^m\dots\sum_{j_n=j_{n-1}}^m\pi(0,j_2,\dots,j_n)B_{0j_2\dots j_n}\nonumber\\
&\langle\psi|\hat I\otimes\bigotimes_{i=2}^n\hat A_{j_i}|\psi\rangle+\nonumber\\
&\eta^n\sum_{j_1=1}^m\sum_{j_2=j_1}^m\dots\sum_{j_n=j_{n-1}}^m\pi(j_1,j_2,\dots,j_n)B_{j_1j_2\dots j_n}\nonumber\\
&\langle\psi|\bigotimes_{i=1}^n\hat A_{j_i}|\psi\rangle>0.
\label{eq:Bellqvioleta}
\end{align}
Then $\eta_{crit}$ is minus one times the ratio of the $(n-1)$-party and $n$-party contributions in the expression above. The former must be negative while the latter positive such that the left hand side is positive above $\eta_{crit}$.

In the next section we will show that with an appropriate choice of the state vector and the measurement settings, the quantum conditional probabilities associated with each Bell coefficient given in Eqs.~(\ref{eq:nm1partycoeff}) and (\ref{eq:npartycoeff}) are equal.
As the same number appears both in the numerator and the denominator of the ratio giving
$\eta_{crit}$, we can simplify with it.
The factor $\pi$ giving the number of independent permutations of the indices is $n!/(n-m+1)!$ for the $(n-1)$-party coefficient and for the $n$-party coefficient given in the first line of
Eq.~(\ref{eq:npartycoeff}). The factor is one half of this for the next $m-2$ coefficients, and
it is  $n!/(n-m+2)!$ for the last one. Taking into account the values of the coefficients we can give the expression for $\eta_{crit}$ as:
\begin{equation}
\eta_{crit}=\frac{1}{1+\frac{m-2}{2}+\frac{n-m+1}{n-m+2}}=\frac{2}{2+m-\frac{2}{n-m+2}}.
\label{eq:etacrit}
\end{equation}
Here we have simplified the fraction by the common factor $n!/(n-m+1)!$. This expression is our main result. Our Bell inequalities are such that their violation, at least in principle, can be detected by using detectors of as low efficiency as given above. Therefore, this is an upper bound for the threshold detector efficiency which may be achieved for $n$-party $m\leq n$ setting binary outcome Bell inequalities. The expression gives $2/(m+1)$ for $n=m$ and
$2/(m+2)$ in the limit of large $n$. However, if we increase $n$, we can achieve less than
by increasing both $m$ and $n$ by just one.

To prove Eq.~(\ref{eq:etacrit}) we must show that the relevant quantum conditional probabilities are really equal as we have claimed, and that the local bound is zero. We will do this in the next two sections.

\section{The quantum conditional probabilities}\label{qcond}
Let all measurements be real ones, that is performed in the $X-Z$
plane. Then each operator $\hat A_j$ can be characterized by a
single real variable $\phi_j$:
\begin{align}
&\hat
A_j|0\rangle=\frac{1-\cos\phi_j}{2}|0\rangle-\frac{\sin\phi_j}{2}|1\rangle
\equiv c_j^-|0\rangle+s_j|1\rangle\nonumber\\
&\hat
A_j|1\rangle=-\frac{\sin\phi_j}{2}|0\rangle+\frac{1+\cos\phi_j}{2}|1\rangle
\equiv s_j|0\rangle+c_j^+|1\rangle. \label{eq:measops}
\end{align}
A zero angle corresponds to the measurement giving outcome one with probability one
for the $|1\rangle$ state. Let all measurement angles be small, and let they obey the following hierarchy:
\begin{equation}
0\leq|\phi_1|\ll|\phi_2|\ll|\phi_3|\ll\dots\ll|\phi_m|.
\label{eq:phihierarchy}
\end{equation}
Let the state be
\begin{equation}
|\psi\rangle=\cos\alpha|{\bf 0}\rangle-\sin\alpha|W\rangle,
\label{eq:psi}
\end{equation}
where
\begin{align}
|{\bf 0}\rangle&\equiv|00\dots 0\rangle\label{eq:0state}\\
|W\rangle&\equiv\frac{1}{\sqrt{n}}(|10\dots 0\rangle+|01\dots 0\rangle+\dots+|00\dots 1\rangle).\label{eq:Wstate}
\end{align}
This state is permutationally symmetric. The angle $\alpha$ will be small, therefore the state is predominantly the separable $|{\bf 0}\rangle$ state with a very small amount of $|W\rangle$ state added.

First we calculate the quantum conditional probability multiplying the $(n-1)$-party Bell coefficient given by Eq.~(\ref{eq:nm1partycoeff}). From Eqs.~(\ref{eq:qvalprob}) and (\ref{eq:psi})
we get:
\begin{align}
&P_Q(A_{10}A_{22}A_{33}\dots A_{mm}A_{(m+1)m}\dots A_{nm})=\nonumber\\
&\cos^2\alpha S_{\bf 00}-2\cos\alpha\sin\alpha S_{W{\bf 0}}+\sin^2\alpha S_{WW},
\label{eq:Pqsubcorr}
\end{align}
where
\begin{equation}
S_{\sigma\tau}\equiv \langle\sigma|\hat I\otimes\bigotimes_{i=2}^n\hat A_{i}|\tau\rangle,
\label{eq:Subcmatet}
\end{equation}
with $\hat A_i\equiv\hat A_m$ when $i\geq m$. As the first step to calculate these matrix elements, from Eqs.~(\ref{eq:measops}) and (\ref{eq:0state}) we get:
\begin{equation}
\hat I\otimes\bigotimes_{i=2}^n\hat A_{i}|{\bf 0}\rangle=|0\rangle\otimes\bigotimes_{i=2}^n(c_i^-|0\rangle+s_i|1\rangle).
\label{eq:subopzero}
\end{equation}
Then, by using Eqs.~(\ref{eq:0state}) and (\ref{eq:Wstate}) it is easy to calculate $S_{\bf 00}$ and $S_{W{\bf 0}}$. We may also get $S_{WW}$ from a similar, but somewhat lengthier calculation.
The result is:
\begin{align}
S_{\bf 00}&=\prod_{i=2}^nc_i^-\nonumber\\
S_{W{\bf 0}}&=\frac{1}{\sqrt{n}}\sum_{k=2}^n\frac{s_k}{c_k^-}\prod_{i=2}^nc_i^-\equiv\frac{S_{\bf 00}}{\xi}\nonumber\\
S_{WW}&=S_{\bf 00}\left(\frac{1}{n}+\frac{1}{\xi^2}\right).
\label{eq:Subcelem}
\end{align}
As $|c_i^-|\ll|s_i|$, $\xi$ is a small number. By substituting Eqs.~(\ref{eq:Subcelem}) into
Eq.~(\ref{eq:Pqsubcorr}) we get
\begin{align}
&P_Q(A_{10}A_{22}A_{33}\dots A_{mm}A_{(m+1)m}\dots A_{nm})=\nonumber\\
&S_{\bf 00}\left[\left(\cos\alpha-\frac{\sin\alpha}{\xi}\right)^2+\frac{\sin^2\alpha}{n}\right].
\label{eq:Pqsubcorrres}
\end{align}
The first term in the square bracket dominates over the second one. However, as we will see later, the contribution from the $n$-party terms is of the same order as the second term. The threshold efficiency may only be finite if the first term vanishes. We can achieve that by choosing the mixing angle $\alpha$ characterizing the quantum state such that ${\rm tg}\alpha=\xi$. With this choice
\begin{equation}
P_Q(A_{10}A_{22}A_{33}\dots A_{mm}A_{(m+1)m}\dots A_{nm})=\frac{S_{\bf 00}\sin^2\alpha}{n}.
\label{eq:Pqsubcorrresu}
\end{equation}

Now let us calculate a general $n$-party conditional probability. Analogously to Eqs.~(\ref{eq:Pqsubcorr}) and (\ref{eq:Subcmatet}) we may write:
\begin{align}
&P_Q(A_{1j_1}A_{2j_2}\dots A_{nj_n})=\nonumber\\
&\cos^2\alpha F_{\bf 00}-2\cos\alpha\sin\alpha F_{W{\bf 0}}+\sin^2\alpha F_{WW},
\label{eq:Pqfcorr}
\end{align}
where
\begin{equation}
F_{\sigma\tau}\equiv \langle\sigma|\bigotimes_{i=1}^n\hat A_{ij_i}|\tau\rangle.
\label{eq:Fcmatet}
\end{equation}
We can calculate these matrix elements similarly to the ones given in Eq.~(\ref{eq:Subcelem}), and we get:
\begin{align}
F_{\bf 00}&=\prod_{i=1}^nc_{j_i}^-\nonumber\\
F_{W{\bf 0}}&=\frac{1}{\sqrt{n}}\sum_{k=1}^n\frac{s_{j_k}}{c_{j_k}^-}\prod_{i=1}^nc_{j_i}^-\equiv\frac{F_{\bf 00}}{\chi}\nonumber\\
F_{WW}&=\frac{F_{\bf 00}}{\chi^2}.
\label{eq:Fulcelem}
\end{align}
By substituting Eqs.~(\ref{eq:Fulcelem}) into
Eq.~(\ref{eq:Pqfcorr}) we get
\begin{align}
P_Q(A_{1j_1}A_{2j_2}\dots A_{nj_n})=&F_{\bf 00}\left(\cos\alpha-\frac{\sin\alpha}{\chi}\right)^2=\nonumber\\
&F_{\bf 00}\sin^2\alpha\left(\frac{1}{\xi}-\frac{1}{\chi}\right)^2.
\label{eq:Pqfc}
\end{align}
Here we used the relation ${\rm tg}\alpha=\xi$.

Let us consider the quantum conditional probabilities multiplying the $n$-party coefficients whose values are given explicitly in Eq.~(\ref{eq:npartycoeff}).
The product giving $F_{\bf 00}$ includes exactly the same elements as the one giving
$S_{\bf 00}$, plus one more (see Eqs.~(\ref{eq:Subcelem}) and (\ref{eq:Fulcelem}). The same is
true for the sums defining $\chi$ and $\xi$. The extra element is the one of index $l$ for
the $l$th line of Eq.~(\ref{eq:npartycoeff}). Then $F_{\bf 00}=S_{\bf 00}c_l^-$ and
$1/\chi=1/\xi+s_l/(c_l^-\sqrt{n})$. By substituting these values into Eq.~(\ref{eq:Pqfc})
we  get that the result is $s_l^2/c_l^-$ times the value given in Eq.~(\ref{eq:Pqsubcorrres})
for the $(n-1)$-party conditional probability. But this extra factor is one in the limit of small angles. Therefore, for our state and measurement settings, the quantum conditional probabilities multiplying each of the Bell coefficients of Eqs.~(\ref{eq:nm1partycoeff})
and (\ref{eq:npartycoeff}) are equal indeed, and have the value given in Eq.~(\ref{eq:Pqsubcorrresu}).

Now, let us consider which are the Bell coefficients (Eq.~\ref{eq:npartycoeffneg}) whose value may freely be chosen. From Eqs.~(\ref{eq:Pqsubcorrresu}) and (\ref{eq:Pqfc}) it follows that the quantum probability characterized by indices $k_1\leq k_2\leq\dots\leq k_n$ is negligible compared to the significant ones if
\begin{equation}
\frac{\prod_{i=1}^nc_{k_i}^-}{\prod_{i=2}^nc_i^-}\left(\sum_{i=2}^n\frac{s_i}{c_i^-}-\sum_{i=1}^n\frac{s_{k_i}}{c_{k_i}^-}\right)^2\ll 1.
\label{eq:negl1}
\end{equation}
Here we substituted the values of $S_{\bf 00}$, $F_{\bf 00}$, $\xi$ and $\chi$ from
Eqs.~(\ref{eq:Subcelem}) and (\ref{eq:Fulcelem}). However,
$c_i^-=(1-\cos\phi_i)/2\approx\phi_i^2/4$ and $s_i=-\sin\phi_i/2\approx-\phi_i/2$, if the angles are small, therefore the above condition may be written as:
\begin{equation}
\left[\frac{\prod_{i=1}^n\phi_{k_i}}{\prod_{i=2}^n\phi_i}\left(\sum_{i=1}^n\frac{1}{\phi_{k_i}}-\sum_{i=2}^n\frac{1}{\phi_i}\right)\right]^2\ll 1.
\label{eq:negl}
\end{equation}
If $k_1=1$, due to the hierarchy imposed by Eq.~(\ref{eq:phihierarchy}) the dominant term in the
parentheses is $1/\phi_1$. Then the condition above may be written as
$(\prod_{i=2}^n\phi_{k_i}/\prod_{i=2}^n\phi_i)^2\ll 1$. This relation holds if
$\phi_{k_i}\leq\phi_i$ for $2\leq i\leq n$, and for at least one index $w$ the inequality is
strict, in which case $\phi_{k_w}\ll\phi_w$. When deriving the formulae we have implicitly supposed that $\phi_1\neq 0$. However, the result is true for $\phi_1=0$ as well. If $k_i=i+1$ for $i<l$, while $k_l=k_{l-1}=l$, the dominant term in the parentheses is $1/\phi_l$ and the condition becomes $(\prod_{i=l+1}^n\phi_{k_i}/\prod_{i=l+1}^n\phi_i)^2\ll 1$. Then if the same relations hold between $\phi_i$ and $\phi_{k_i}$ as for the $k_1=1$ case, but now for
$l+1\leq i\leq n$, Eq.~(\ref{eq:negl}) will hold and the threshold efficiency will be independent of the choice of the associated Bell coefficient.

\section{The classical case}\label{class}
For an $n$-party Bell inequalitiy with $m$ measurement settings
per party the classical bound is zero if
\begin{equation}
\sum_{j_1=0}^m\sum_{j_2=0}^m\dots\sum_{j_n=0}^m  B_{j_1j_2\dots j_n}
a_{1j_1}a_{2j_2}\dots a_{nj_n}\leq 0,
\label{eq:clbound0gen}
\end{equation}
for all deterministic strategies, which are defined by the actual choices of
$a_{ij}$ ($i=1,\dots,n$; $j=1,\dots,m$), where each of them may take the value of zero or one,
while $a_{i0}=1$ for all $i$. The $a_{i0}$ appear in the subcorrelation terms. A deterministic strategy means that each party has a definite outcome for each measurement setting
with probability one. This outcome for the $j$th measurement of the $i$th party is given by $a_{ij}$. Let us denote the matrix with elements $a_{ij}$ by $\bar a$. Then for permutationally
symmetric inequalities Eq.~(\ref{eq:clbound0gen}) may be rewritten as:
\begin{equation}
\sum_{j_1=0}^{j_2}\sum_{j_2=0}^{j_3}\dots\sum_{j_n=0}^m  B_{j_1j_2\dots j_n}
C(\bar a,j_1,j_2,\dots,j_n)\leq 0,
\label{eq:clbound0per}
\end{equation}
where
\begin{align}
&C(\bar a,j_1,j_2,\dots,j_n)=
\frac{\pi(j_1,j_2,\dots,j_n)}{n!}\sum_{\sigma\in S_n}\prod_{i=1}^n a_{ij_{\sigma(i)}}=
\nonumber\\
&\frac{\pi(j_1,j_2,\dots,j_n)}{n!}{\rm perm}(\bar\alpha(\bar a,j_1,j_2,\dots,j_n)).
\label{eq:Cadef}
\end{align}
The sum in the first line extends over all elements $\sigma$ of the symmetric group $S_n$, that is over all
permutations of the numbers $1,2,\dots n$. We should only have summed over the independent permutations of the numbers $j_1,j_2\dots j_n$. The difference is taken care of by the factor
in front of the summation with $\pi(j_1,j_2,\dots,j_n)$ denoting the number of independent permutations of its arguments. The sum itself is nothing else but the permanent of the
$n\times n$ square matrix $\bar\alpha(\bar a,j_1,j_2,\dots,j_n)$ consisting of the $j_1$th, $j_2$th, ..., $j_n$th columns of the $\bar a$ strategy matrix.

For the Bell inequality given by Eqs.~(\ref{eq:nm1partycoeff})-(\ref{eq:npartycoeffneg}) the condition Eq.~(\ref{eq:clbound0per}) that the classical bound is zero may be written as:
\begin{align}
&2{\rm perm}(\bar\alpha(\bar a,1234\dots mm\dots m))+\nonumber\\
&{\rm perm}(\bar\alpha(\bar a,2234\dots mm\dots m))+\nonumber\\
&{\rm perm}(\bar\alpha(\bar a,2334\dots mm\dots m))+\nonumber\\
&{\rm perm}(\bar\alpha(\bar a,2344\dots mm\dots m))+\dots+\nonumber\\
&2\frac{n-m+1}{n-m+2}{\rm perm}(\bar\alpha(\bar a,234\dots mmm\dots m))-\nonumber\\
&2{\rm perm}(\bar\alpha(\bar a,0234\dots mm\dots m))-Q(\bar a)\leq 0.
\label{eq:clbound0our}
\end{align}
Here we have used Eq.~(\ref{eq:Cadef}), we have substituted the actual values of
$\pi(j_1,j_2,\dots,j_n)$ and simplified the inequality by a factor of $2(n-m+1)!$.
The $Q(\bar a)\geq 0$ denotes the contribution of terms due to the negative $n$-party Bell coefficients of Eq.~(\ref{eq:npartycoeffneg}), multiplied by minus $2(n-m+1)!$.

A permanent is independent of the order of the rows and columns of the matrix. Therefore,
may rearrange the columns of the matrices appearing in Eq.~(\ref{eq:clbound0our})
such that their $j$th column is the $(j+1)$th column of $\bar a$ if $j\leq m-1$,
the $m$th column if $m-1\leq j\leq n-1$ and only their last column differ from
each other. However, the sum of two permanents whose matrices differ from each other only in one single column (row) is the permanent of the matrix having the sum of the
those columns (rows) in that position. Furthermore, multiplying a permanent with a factor has the same effect as multiplying one of the columns (rows) of the matrix.
Therefore,  Eq.~(\ref{eq:clbound0our}) may be rewritten as:
\begin{equation}
{\rm perm}(\bar\beta(\bar a))-Q(\bar a)\leq 0,
\label{eq:clbound0oursimp}
\end{equation}
where
\begin{align}
&\beta_{ij}=a_{i(j+1)}\quad\quad\quad\quad\quad\quad\quad j\leq m-1\nonumber\\
&\beta_{ij}=a_{im}\quad\quad\quad\quad\quad\quad m-1\leq j\leq n-1\nonumber\\
&\beta_{in}=2a_{i1}+\sum_{k=2}^{m-1}a_{ik}+2\frac{n-m+1}{n-m+2}a_{im}-2.
\label{eq:betamat}
\end{align}

Our Bell inequality is appropriate if Eq.~(\ref{eq:clbound0our}) holds for all deterministic strategies. The permanents appearing in the expression can not be negative, as all the elements of the matrices involved are zero or one. Therefore, strategies giving zero for all terms with a positive sign trivially satisfy
Eq.~(\ref{eq:clbound0our}). We also do not have to consider explicitly any strategy leading to a
nonzero factor for a Bell coefficient whose value does not affect the threshold efficiency
(Eq.~\ref{eq:npartycoeffneg}). The $Q(\bar a)$ depends linearly on such a coefficient, if we choose its absolute value large enough, we can always satisfy Eq.~(\ref{eq:clbound0our}).

First let us consider strategies giving a positive value for the first term of
Eq.~(\ref{eq:clbound0our}). This is true if and only if there is at least one party having outcome one for the first measurement setting, at least one other party having outcome one for the second measurement setting, and so on, and finally, there are at least $n-m+1$ other parties having outcome one for their $m$th measurement setting. For a permutationally invariant inequality all strategies differing from each other only in the order of the parties give the same constraints for the Bell coefficients. Therefore, without sacrificing generality, we may assume that the first party's first measurement outcome is one, the second
party's second measurement outcome is one, and so on, and finally, the last $n-m$ party's $m$th measurement outcome is one, that is $a_{ik}=1$ where $k={\rm min}(i,m)$. We only have to consider strategies whose matrices contain zeros to the left of these elements in each row,
that is $a_{ik'}=0$ if $0<k'<k$. If $a_{ik'}$ were one, then
the factor of $B_{12\dots(i-1)k'(i+1)\dots mm\dots}$ would not be  zero.
However, the quantum conditional probability associated with this coefficient is
$(\phi_{k'}/\phi_k)^2\ll 1$ times the one associated with the significant coefficients (this follows from Eq.~(\ref{eq:negl}), therefore, we need not consider such a strategy explicitly indeed.
The matrix of the remaining strategies for $m=6$ and $n=9$, without the zeroth column is:
\begin{equation}
\left(\begin{array}{cccccc}
1&a_{12}&a_{13}&a_{14}&a_{15}&a_{16}\\
0&1&a_{23}&a_{24}&a_{25}&a_{26}\\
0&0&1&a_{34}&a_{35}&a_{36}\\
0&0&0&1&a_{45}&a_{46}\\
0&0&0&0&1&a_{56}\\
0&0&0&0&0&1\\
0&0&0&0&0&1\\
0&0&0&0&0&1\\
0&0&0&0&0&1\\
\end{array}\right).
\label{eq:69strat}
\end{equation}
With such a strategy matrix $Q(\bar a)$ appearing in Eqs.~(\ref{eq:clbound0our}) and
(\ref{eq:clbound0oursimp}) is zero: terms contributing to it have zero factors. It is also clear that increasing the number of parties beyond $m$ does not allow any more free parameters.

For this example Eq.~(\ref{eq:clbound0oursimp}) may be written as:
\begin{equation}
{\rm perm}\left(\begin{array}{ccccccccc}
a_{12}&a_{13}&a_{14}&a_{15}&a_{16}&a_{16}&a_{16}&a_{16}&\beta_{1n}\\
1&a_{23}&a_{24}&a_{25}&a_{26}&a_{26}&a_{26}&a_{26}&\beta_{2n}\\
0&1&a_{34}&a_{35}&a_{36}&a_{36}&a_{36}&a_{36}&\beta_{3n}\\
0&0&1&a_{45}&a_{46}&a_{46}&a_{46}&a_{46}&\beta_{4n}\\
0&0&0&1&a_{56}&a_{56}&a_{56}&a_{56}&\beta_{5n}\\
0&0&0&0&1&1&1&1&\beta_{mn}\\
0&0&0&0&1&1&1&1&\beta_{mn}\\
0&0&0&0&1&1&1&1&\beta_{mn}\\
0&0&0&0&1&1&1&1&\beta_{mn}\\
\end{array}\right)\leq 0.
\end{equation}
Let us expand the permanent along its last row. There are
$n-m+1$ ones in that row, all in columns equal to each other. These terms of the expansion
give $n-m+1$ times the permanent of the matrix with that colunm and the last row deleted.
The contribution from the last element will be $\beta_{mn}$ times the permanent of the matrix with its last row and last column deleted. The sum of the contributions may be expressed as
the permanent of a single matrix whose $i$th element in its last column is
$(n-m+1)\beta_{in}+\beta_{mn}a_{im}$, which we can calculate from Eqs.~(\ref{eq:betamat}) and (\ref{eq:69strat}). The value we get is
\begin{equation}
(n-m+1)\left(2a_{i1}+\sum_{k=2}^{m-1}a_{ik}+2\frac{n-m}{n-m+1}a_{im}-2\right),
\end{equation}
which is nothing else that
$n-m+1$ times the last column of the matrix with $n-1$ parties. Therefore, the value of the permanent
for $n$ parties is $n-m+1$ times the value for $n-1$ parties. If we repeat this $n-m$ times we
get $(n-m+1)!$ times the permanent for $n=m$.

Let us consider $n=m$. For $m=6$ Eq.~(\ref{eq:clbound0oursimp}) may be written as:
\begin{equation}
{\rm perm}\left(\begin{array}{cccccl}
a_{12}&a_{13}&a_{14}&a_{15}&a_{16}&\sum_{j=2}^6a_{1i}\\
1&a_{23}&a_{24}&a_{25}&a_{26}&\sum_{j=3}^6a_{2i}-1\\
0&1&a_{34}&a_{35}&a_{36}&\sum_{j=4}^6a_{3i}-1\\
0&0&1&a_{45}&a_{46}&a_{45}+a_{46}-1\\
0&0&0&1&a_{56}&a_{56}-1\\
0&0&0&0&1&-1\\
\end{array}\right)\leq 0.
\label{eq:betmn6}
\end{equation}
The values in the last column follow from Eq.~(\ref{eq:betamat}). If we expand the permanent along its
last row we get the permanent of a matrix of the same form corresponding to $m-1$ measurement settings.
If we repeat this $m-1$ times, finally we get zero for the permanent. Therefore,
Eq.~(\ref{eq:clbound0oursimp}) and equivalently, Eq.~(\ref{eq:clbound0our}) is satisfied as an
equality. Although we demonstrated this result on the example of $m=6$, it is obvious that it is
true for any $m$.

So far we have only dealt with strategies that lead to a nonzero factor for the first term of
Eq.~(\ref{eq:clbound0our}). Now let us consider strategies giving zero for the first
$l-1$ terms of Eq.~(\ref{eq:clbound0our}) and nonzero for the $l$th term. If
$n=m=8$ and $l=5$ the strategy matrix to be considered without the zeroth column is the following:
\begin{equation}
\left(\begin{array}{cccccccc}
a_{11}&1&a_{13}&a_{14}&a_{15}&a_{16}&a_{17}&a_{18}\\
a_{21}&a_{22}&1&a_{24}&a_{25}&a_{26}&a_{27}&a_{28}\\
a_{31}&a_{32}&a_{33}&1&a_{35}&a_{36}&a_{37}&a_{38}\\
0&0&0&0&1&a_{46}&a_{47}&a_{48}\\
0&0&0&0&1&a_{56}&a_{57}&a_{58}\\
0&0&0&0&0&1&a_{67}&a_{68}\\
0&0&0&0&0&0&1&a_{78}\\
0&0&0&0&0&0&0&1\\
\end{array}\right).
\end{equation}
Elements $a_{k(l-1)}=0$ and $a_{kl}=0$ for $k<l$, otherwise the $k$th term of
Eq.~(\ref{eq:clbound0our}) would be positive. Furthermore, $a_{ij}=0$ if $j<i$ for $i>l$.
For example, $a_{73}$ above is zero. If it were not, the factor multiplying
$B_{23345568}$ would be positive. The quantum conditional probability associated to this coefficient is small by a factor of $(\phi_5/\phi_7)^2\ll 1$ (see Eq.~(\ref{eq:negl})). For the interesting strategies a larger number of parties would mean an additional $n-m$ copies of the last line, which contains a single one in the last position. Only this property of the matrix was used when we showed that increasing $n$ beyond $m$ only introduces a positive factor, which does not affect the validity of the inequality. Therefore, it is enough to deal with $n=m$.

For the present example Eq.~(\ref{eq:clbound0oursimp}) may be written as:
\begin{equation}
{\rm perm}\left(\begin{array}{cccccccc}
1&a_{13}&a_{14}&a_{15}&a_{16}&a_{17}&a_{18}&\beta_{1m}\\
a_{22}&1&a_{24}&a_{25}&a_{26}&a_{27}&a_{28}&\beta_{2m}\\
a_{32}&a_{33}&1&a_{35}&a_{36}&a_{37}&a_{38}&\beta_{3m}\\
0&0&0&1&a_{46}&a_{47}&a_{48}&\beta_{4m}\\
0&0&0&1&a_{56}&a_{57}&a_{58}&\beta_{5m}\\
0&0&0&0&1&a_{67}&a_{68}&\beta_{6m}\\
0&0&0&0&0&1&a_{78}&\beta_{7m}\\
0&0&0&0&0&0&1&\beta_{8m}\\
\end{array}\right)\leq 0.
\end{equation}
This permanent is equal to the product of the permanents of two blocks along the diagonal of sizes $(l-2)\times(l-2)$ and $(n-l+2)\times(n-l+2)$. In the present case
\begin{equation}
{\rm perm}\left(\begin{array}{ccc}
1&a_{13}&a_{14}\\
a_{22}&1&a_{24}\\
a_{32}&a_{33}&1\\
\end{array}\right)\times
{\rm perm}\left(\begin{array}{ccccc}
1&a_{46}&a_{47}&a_{48}&\beta_{4m}\\
1&a_{56}&a_{57}&a_{58}&\beta_{4m}\\
0&1&a_{67}&a_{68}&\beta_{4m}\\
0&0&1&a_{78}&\beta_{4m}\\
0&0&0&1&\beta_{4m}\\
\end{array}\right).
\label{eq:twopermsum}
\end{equation}
If we substitute the values of the last column of the second matrix from Eq.~(\ref{eq:betamat}),
we get a matrix whose structure is very similar to the one in Eq.~(\ref{eq:betmn6}).
There is one important difference. Now the last elements of each row is always the sum of the preceding elements minus two. In the case of Eq.~(\ref{eq:betmn6}) this is not true for the first row, the
minus two is missing. Due to this difference, if we follow the steps we have taken in the case of Eq.~(\ref{eq:betamat}), we will get not zero, but minus two. Therefore, Eq.~(\ref{eq:clbound0our})
is satisfied now as a strict inequality.

\section{The 333 case}
\label{333}

For $m=n=3$ a permutationally invariant inequality with the properties required may be defined
by the following coefficients:
\begin{align}
&B_{023}=-1 && B_{112}=-1 && B_{113}=-1 && B_{122}=-2 && \nonumber\\
&B_{123}=+1 && B_{223}=+1 && B_{233}=+1. \label{eq:B333}
\end{align}
Bell coefficients having the same set of indices in different orders as any of the coefficients above have the same value. All other coefficients are zero.

\begin{figure}[ht!]
\vspace{0cm}
\centering
\includegraphics[angle=-90, width=\columnwidth]{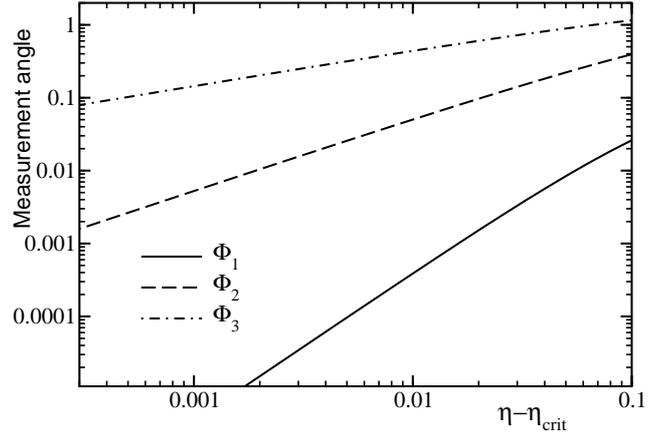}
\caption{The optimum measurement angles for the $m=n=3$ inequality as functions of the detector efficiency} \label{fig:angles}
\end{figure}

We have got this inequality by iterating a linear programming step
and a semidefinite programming one. The former one provides the
Bell inequality of zero classical bound with no single-party
marginals having the lowest critical detector efficiency from
known matrix elements of the measurement operators. The procedure
is the same as the one we used in Ref.~\cite{PVw}. In the
semidefinite programming step we applied the method of
Navascu\'es, Pironio and Ac\'\i n \cite{NPA} at level 3 (that is,
the maximum length of all tuples of operators is 3 in the
sequence) to get an approximation for the matrix elements and a
quite tight upper bound for the violation. As starting values, we
have chosen three random settings on the Bloch sphere (the same
for each party) along with a random symmetric 3-qubit state. Note
the initial qubit measurements do not confine in general the state
to a 3-qubit state during the iteration procedure. We have
repeated the above iterative procedure many times with different
(randomly generated) starting values. The best inequality we have
got is the one given above. We note that the only assumption we
have made is that we allowed no single-party marginals, we
actually allowed non-symmetric inequalities, and the procedure
constrained neither the state, nor the measurement operators, not
even the dimensionality of the Hilbert space.

\begin{figure}[ht!]
\vspace{1cm}
\centering
\includegraphics[angle=-90, width=\columnwidth]{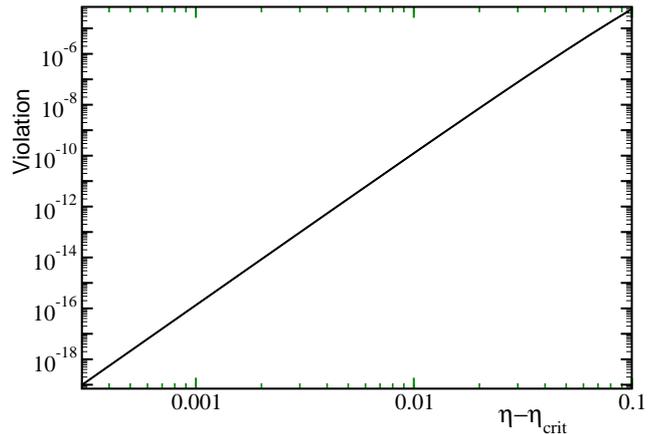} \caption{The maximum violation of the $m=n=3$ inequality as a function of the detector efficiency} \label{fig:viola}
\end{figure}

We applied the see-saw algorithm \cite{seesaw} to the inequality using three-qubit states. From this calculation it turned out that this space was enough to get the violation and $\eta_{crit}=1/2$. We could also find out from this calculation the structure of the state and the behavior of the measurement operators. We have also derived a Bell inequality with $m=3$ and $n=4$, and we have studied its behavior with the see-saw algorithms. These results, and the ones for $m=2$ and $n>2$ by Larsson and Semitecolos \cite{LS} made it possible for us to make the generalization and to get the family of inequalities presented in this paper.

We have calculated numerically the optimum measurement angles and the maximum violation as functions of the detector efficiency above the threshold efficiency
$\eta_{crit}=0.5$ for the inequality given in Eq.~(\ref{eq:B333}). To get more accurate numbers, we
have not used the see-saw algorithm to get these results, only the knowledge we had gained from it about the form of the solution. We have used the analytical expressions given by Eqs.~(\ref{eq:Subcelem}) and (\ref{eq:Fulcelem}) for the matrix elements, and optimized numerically the measurement angles and the mixing angle $\alpha$.

The results are shown in Figs.~\ref{fig:angles} and \ref{fig:viola}. The first measurement angle tends to zero at the fastest pace, it behaves as the square of $\eta-\eta_{crit}$. If we choose its value exactly zero, the violation changes very little. The second measurement angle is proportional with $\eta-\eta_{crit}$, while the third one behaves as
the square root of $\eta-\eta_{crit}$. The violation is is proportional with the sixth power
of $\eta-\eta_{crit}$.

\section{Summary}
In the present paper we have defined a family of binary outcome $n$ party Bell inequalities with $m\leq n$ measurement settings for each party, whose $m>2$ members can be violated by less detection efficiency than any other inequalities known so far. This gives upper bounds for the minimum value $\eta_*$ of the critical efficiency achievable for such inequalities. The family is the generalization of the one given by Larsson and Semitecolos \cite{LS} for $m=2$.  Unfortunately, for $m>2$ there is still a gap between our upper bounds and the best lower bounds $n/((n-1)m+1)$ given by Massar and Pironio \cite{massar}. For $m=n=3$
our upper bound is $1/2$, while the lower bound is $3/7$. As we have found no better Bell inequality while making assumptions neither about the symmetry nor about the violating state, $\eta_*$ may well agree with our bound for both this case, and also for $m=3$; $n=4$.
For $n=m$ our bound Eq.~(\ref{eq:etacrit}) gives $\eta_*\leq 2/(m+1)$, while from Ref.~\cite{massar} $\eta_*\geq 1/(m-1+1/m)$ follows, a factor of two difference for large $m$. Buhrman \etal \cite{BHMR} gives a worse upper bound of $\eta_*\leq 8/m$. When $n\rightarrow\infty$, $\eta_*\leq 2/(m+2)$, as it follows from Eq.~(\ref{eq:etacrit}). In this case the lower bound is $\eta_*\geq 1/m$, so there is still a factor of two difference for large $m$. The upper bound of Ref.~\cite{BHMR} behaves very similarly to ours, it is $\eta_*\geq 1/m$ (see Eq.~(4) in their paper). It is interesting to note that $\eta_*$ does not approach zero if $n$ alone goes to infinity, it is proportional to $1/m$. However, if $m\rightarrow\infty$ while $n$ remains finite, the lower bound of Ref.~\cite{massar} goes to zero. Ref.~\cite{BHMR} does not give a useful upper bound for this situation, while in the case of Ref.~\cite{PVB} the critical efficiency remains finite, proportional to $1/n$. Unfortunately, the present work tells nothing about the $m>n$ case, so even the qualitative behavior of $\eta_*$ in this case remains open.

For the present family of Bell inequalities the violating state
approaches a product state near the threshold efficiency. The same
behavior has been observed by Vallone \etal \cite{Vallonetal} for
several other Bell inequalities, e.~g.\ for the two-qubit chained
inequalities with any number of measurement settings. They even
concluded that the complementary behavior of entanglement and
nonlocality might be general. In the multipartite case this can
not be so, as the inequalities given in Refs.~\cite{PVB} and
\cite{PVw} are violated by the maximally entangled GHZ and $W$
states near $\eta_{crit}$, respectively. In the bipartite case, a
counterexample is shown in the Appendix. However, it is still
possible that the conjecture by Vallone \etal is correct for
inequalities with the smallest critical efficiency, the ones with
$\eta_{crit}=\eta_{*}$.

Our multipartite Bell inequalities involve high-order correlators
among the particles, which are challenging to access in case of
many particles. However, recently experimentally more friendly
binary-outcome Bell inequalities have been constructed involving
only two-body correlators~\cite{twobody}. We pose it as an
interesting problem to find such low-order Bell inequalities which
are suited to Bell violation using detectors of low efficiencies.

\section{Acknowledgements}
We acknowledge financial support from the Hungarian
National Research Fund OTKA (K111734) and a J\'anos Bolyai Grant of
the Hungarian Academy of Sciences.

\appendix
\section{Two-party Bell inequality where the maximal robustness of nonlocality is attained with a maximally entangled state}

To prove the title of the appendix, let us consider the following
Bell inequality:
\begin{align}\label{chtype}
I_{2}\equiv& P(A_1,B_1) + P(A_1,B_2) + P(A_2,B_1) - P(A_2,B_2)\nonumber\\
&- c P(A_1) - c P(B_1)\leq 2-2c,
\end{align}
where $c=\frac{2(2^{1/4}-1)}{\sqrt{2}-1}\simeq0.9136$.
We choose that both Alice and Bob always output ``0" in case of non-detection, hence all measurements have binary outputs (it turns out that other choices will not lead to lower detection efficiency thresholds). Then, the measurement probabilities are modified according to $P(A_x,B_y)\rightarrow\eta^2 P(A_x,B_y)$, $P(A_x)\rightarrow\eta P(A_x)$ and similarly for Bob.
Introducing these expressions in Eq.~(\ref{chtype}) and dividing by $\eta^2$, we obtain the detection-efficiency-dependent inequality:
\begin{align}
\label{CHtype}
I_{2}(\eta) =& P(A_1,B_1) + P(A_1,B_2) + P(A_2,B_1) - P(A_2,B_2)\nonumber \\
&- c\frac{P(A_1)}{\eta} - c\frac{P(B_1)}{\eta}\leq \frac{2 -2c}{\eta^2}.
\end{align}
By setting $\eta=c$, we get the Clauser-Horne inequality on the left-hand-side, whereas the right-hand-side becomes $(\sqrt 2-1)/2$, which is just the maximal quantum violation of the Clauser-Horne inequality attainable with a pair of maximally entangled qubits~\cite{tsirelson}. From this it follows that for any $\eta$ approaching $c\simeq0.9136$ from above the maximal violation of the Bell inequality~(\ref{chtype}) is given by a state converging to the maximally entangled two-qubit state.

\end{document}